\documentclass[10pt,conference]{IEEEtran}
\IEEEoverridecommandlockouts
\usepackage{multirow}
\usepackage{epsfig}
\usepackage{listings}
\usepackage{xcolor}
\usepackage{colortbl}
\usepackage{booktabs}
\usepackage{caption}
\usepackage{subcaption}
\usepackage{diagbox}
\usepackage[hidelinks]{hyperref}
\usepackage[nobreak,space,compress]{cite}
\usepackage{flushend}
\usepackage{amsmath}
\definecolor{codegreen}{rgb}{0,0.6,0}
\definecolor{codegray}{rgb}{0.5,0.5,0.5}
\definecolor{codepurple}{rgb}{0.58,0,0.82}
\definecolor{backcolour}{rgb}{0.95,0.95,0.92}
\usepackage{color}
\definecolor{gray}{rgb}{0.4,0.4,0.4}
\definecolor{darkblue}{rgb}{0.0,0.0,0.6}
\definecolor{cyan}{rgb}{0.0,0.6,0.6}

\lstdefinestyle{JavaStyle}{
    backgroundcolor=\color{backcolour},   
    commentstyle=\color{codegreen},
    keywordstyle=\color{magenta},
    numberstyle=\tiny\color{codegray},
    stringstyle=\color{codepurple},
    basicstyle=\ttfamily\footnotesize,
    breakatwhitespace=false,         
    breaklines=true,                 
    captionpos=b,                    
    keepspaces=true,                 
    numbers=left,                    
    numbersep=5pt,                  
    showspaces=false,                
    showstringspaces=false,
    showtabs=false,                  
    tabsize=2
}

\newcommand{\fastr}{\mbox{FAST-R}}
\newcommand{\ourapproach}{\mbox{ATM}}
\usepackage[hang,flushmargin]{footmisc} 
\newcommand{\summarybox}[1]{
    \vspace{3mm}
    \noindent 
    \framebox[\linewidth][c]{\parbox[b]{0.95\linewidth}{{#1}}}
}

\def\BibTeX{{\rm B\kern-.05em{\sc i\kern-.025em b}\kern-.08em
    T\kern-.1667em\lower.7ex\hbox{E}\kern-.125emX}}

\begin{document}

\title{
{\ourapproach: Black-box Test Case Minimization based on Test Code Similarity and Evolutionary Search
}
\thanks{
This work was supported by a research grant from Huawei Technologies Canada Co., Ltd, as well as by the Mitacs Accelerate Program, the Canada Research Chair and Discovery Grant programs of the Natural Sciences and Engineering Research Council of Canada (NSERC). The experiments conducted in this work were enabled in part by support provided by the Digital Research Alliance of Canada (https://alliancecan.ca).
}
}

\author
{
  \IEEEauthorblockN{\textsf{Rongqi Pan}}
  \IEEEauthorblockA{School of EECS\\University of Ottawa\\Ottawa, Canada\\
  \textsf{rpan099@uottawa.ca}
  }\vspace{-30pt}
\and

  \IEEEauthorblockN{\textsf{Taher A. Ghaleb}}
  \IEEEauthorblockA{School of EECS\\University of Ottawa\\Ottawa, Canada\\
  \textsf{tghaleb@uottawa.ca}
  }\vspace{-30pt}
\and
  \IEEEauthorblockN{\textsf{Lionel Briand}}
  \IEEEauthorblockA{School of EECS, University of Ottawa\\Ottawa, Canada\\SnT Centre, University of Luxembourg\\Luxembourg\\
  \textsf{lbriand@uottawa.ca}
  }
  \vspace{-30pt}
}

\maketitle

\begin{abstract}
Executing large test suites is time and resource consuming, sometimes impossible, and such test suites typically contain many redundant test cases.
Hence, test case (suite) minimization is used to remove redundant test cases that are unlikely to detect new faults.
However, most test case minimization techniques rely on code coverage (white-box), model-based features, or requirements specifications, which are not always (entirely) accessible by test engineers. Code coverage analysis also leads to scalability issues, especially when applied to large industrial systems.
Recently, a set of novel techniques was proposed, called \fastr, relying solely on test case code for test case minimization, which appeared to be much more efficient than white-box techniques. However, it achieved a comparable low fault detection capability for Java projects, thus making its application challenging in practice.
In this paper, we propose \ourapproach~(AST-based Test case Minimizer), a similarity-based, search-based test case minimization technique, taking a specific budget as input, that also relies exclusively on the source code of test cases but attempts to achieve higher fault detection through finer-grained similarity analysis and a dedicated search algorithm.
\ourapproach~transforms test case code into Abstract Syntax Trees (AST) and relies on four tree-based similarity measures to apply evolutionary search, specifically genetic algorithms, to minimize test cases.
We evaluated the effectiveness and efficiency of \ourapproach~on a large dataset of $16$ Java projects with $661$ faulty versions using three budgets ranging from $25\%$ to $75\%$ of test suites.
\ourapproach~achieved significantly higher fault detection rates ($0.82$ on average), compared to \fastr~($0.61$ on average) and random minimization ($0.52$ on average), when running only $50\%$ of the test cases, within practically acceptable time ($1.1-4.3$ hours, on average, per project version), given that minimization is only occasionally applied when many new test cases are created (major releases). Results achieved for other budgets were consistent. 
\end{abstract}


\vspace{-3pt}
\section{Introduction}
Software testing is a widely used verification mechanism to detect faults in software releases.
However, test suites tend to grow in size as software evolves, making the execution of all test cases time and resource consuming~\cite{yoo2012regression}, if not infeasible. Such test suites are prone to similar and redundant test cases that are unlikely to detect different faults and, if not removed, are repeatedly executed many times on many versions, such as in continuous integration contexts. This can lead to a massive waste of time and resources~\cite{yoo2012regression}, especially for large industrial systems, thus warranting systematic and automated strategies to eliminate redundant test cases, known as test case (suite) minimization.

Though many test case minimization techniques exist~\cite{khan2018systematic}, most of them rely on analyzing the test coverage of the system production code (white-box), model-based features, or requirements specifications. Despite their benefits in minimizing test suites, such information is not always (entirely) accessible or available to test engineers, making them not easy to apply in practice. Further, analyzing production code entails many scalability and practicality issues, especially when applied to large industrial systems~\cite{elbaum2014techniques,herzig2018testing}.
One exception is the recent and novel work of Cruciani et al.~\cite{cruciani2019scalable}, called \fastr, that relies solely on the source code of test cases. Though much more efficient than white-box techniques, \fastr~achieved a comparable low fault detection capability for Java test cases.
Unlike test case selection and prioritization, test case minimization is typically performed on an occasional basis~\cite{noemmer2020evaluation}, that is not for every code change but rather at certain milestones, such as major releases when many new test cases are created.
Therefore, a technique that is more time-consuming than \fastr~but runs within practical time and achieves higher fault detection rates would often be a better trade-off in practice.

In this paper, we propose \ourapproach~(\textbf{A}ST-based \textbf{T}est case \textbf{M}inimizer), a test case minimization technique based on tree-based test code similarity and evolutionary search. \ourapproach~is black-box as it relies exclusively on test code, thus requiring no access to the production code of the system under test.  \ourapproach~achieves significantly higher fault detection rates than \fastr~and runs within practical time through a finer-grained analysis of test cases and search-based optimization.
\ourapproach~pre-processes and normalizes test case code,
transforms it into Abstract Syntax Trees (AST), and then compares test case ASTs using four tree-based similarity measures: top-down, bottom-up, combined (merging the first two), and tree edit distance.
Finally, \ourapproach~employs
Genetic Algorithm (GA) and its multi-objective counterpart, Non-Dominated Sorting Genetic Algorithm II (NSGA-II), to minimize test cases using the above similarity measures as fitness, individually or combined.

We evaluated \ourapproach~compared to baseline techniques: \fastr~and random minimization (a standard baseline). In contrast to Cruciani et al.~\cite{cruciani2019scalable}, our evaluation was performed on a large set of Java test cases, extracted from $16$ Java projects with many versions, each of which with a single real fault associated with one or more test case failures. Moreover, while \fastr~was evaluated at the level of Java test classes, our evaluation is finer-grained as it focuses on Java test methods, where each method is considered a test case. This was motivated by the fact that a fault may be detected by only a subset of test methods rather than a whole test class. Therefore, performing minimization at the method level enables the removal of unnecessary test cases in a more precise manner~\cite{vasic2017file,gligoric2015practical}, which has been shown to achieve better results than those at the class level~\cite{zhang2013bridging,mei2012static,de2017test}.
We used the Fault Detection Rate (\textit{FDR}) and execution time as metrics to respectively evaluate the effectiveness and efficiency of \ourapproach~using three minimization budgets, ranging from $25\%$ to $75\%$, covering most of the practical budget range as test engineers usually want to preserve significant test suite fault detection power. We compared the results of all alternative \ourapproach~configurations among themselves and, by also accounting for the time of similarity calculations, we identified the best one and compared it to baseline techniques.
Specifically, we addressed the following research questions.

\begin{itemize}
    \item \textit{RQ1: How does \ourapproach~perform under different configurations in terms of test case minimization?}
    
    For a $50\%$ minimization budget, \ourapproach~achieved high fault detection rates (0.82 on average) and ran within practically acceptable time ($1.1-4.3$ hours on average across configurations), with combined similarity using GA being the best configuration when considering both effectiveness ($0.80$ \textit{FDR}) and efficiency ($1.2$ hours). Such results were consistent for other budgets ($25\%$ and $75\%$). 
    
    \item \textit{RQ2: How does \ourapproach~compare to state-of-the-art black-box test case minimization techniques?}
    
    The best configuration of \ourapproach~outperformed other techniques in terms of effectiveness, by achieving significantly higher \textit{FDR} results than \fastr~($+0.19$ on average) and random minimization ($+0.28$ on average), while running within practically acceptable (though longer) time ($1.2$ hours on average).
\end{itemize}

Overall, this paper makes the following contributions.
\begin{itemize}
    \item A black-box, AST-similarity- and search-based test case minimization technique, called \ourapproach, that offers a better trade-off between effectiveness and efficiency than existing work, in many practical contexts. This includes (a) a finer-grained technique that considers test cases to be Java test methods, pre-processes them, and transforms their code into ASTs; (b) a tree-based similarity measure that merges two complementary similarity measures that have not been used for test case similarity, thus capturing more information about test case commonalities.
    \item A large-scale test case minimization experiment on $16$ projects with $661$ versions comparing several configurations of \ourapproach~and baseline techniques, taking approximately three months of calendar time and $23$ years of computation time on a cluster of $1,304$ nodes with $80,912$ available CPU cores. This is the largest test case minimization experiment to date in the research literature.
\end{itemize}

The rest of this paper is organized as follows.
Section~\ref{Approach} presents our proposed technique for test case minimization.
Section~\ref{Validation} presents the experiment design, reports experimental results, and discusses their practical implications.
Section~\ref{Threats} discusses the validity threats to our results. 
Section~\ref{RelatedWork} reviews and contrasts our technique with related work.
Finally, Section~\ref{Conclusion} concludes the paper and suggests future work.

\section{\ourapproach: Black-box Test Case Minimization}
\label{Approach}

\begin{figure*}[ht]
    \centering
    \includegraphics[width=.9\textwidth]{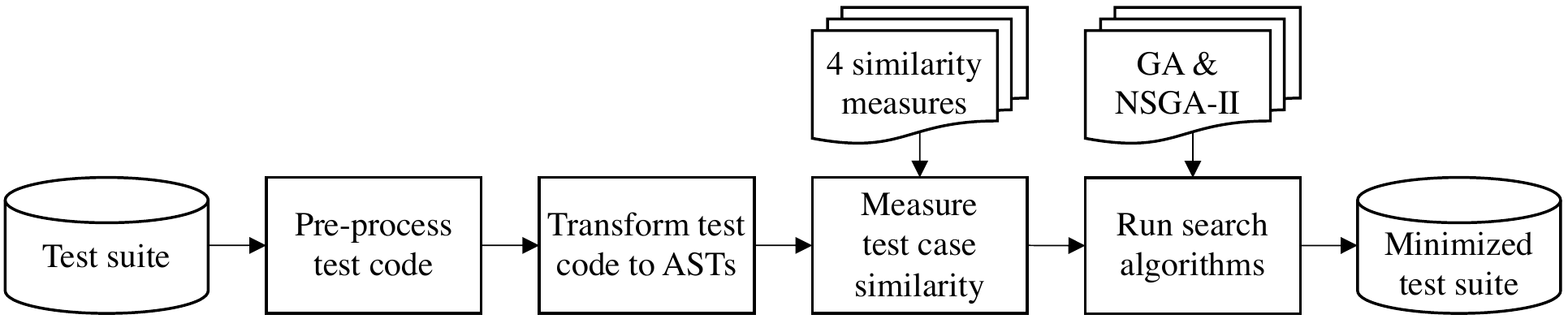}
    \caption{An overview of the main steps of \ourapproach~to perform test case minimization}
    \label{fig:approach}
    \vspace{-4pt}
\end{figure*}

This section describes our black-box technique, called \ourapproach, for test case minimization relying on tree-based test code similarity and evolutionary search.
Figure~\ref{fig:approach} gives an overview of the main steps of \ourapproach.
We first describe how we pre-process the source code of test cases (Section~\ref{preprocessing}) and transform them into ASTs (Section~\ref{CodeAsAST}). Then, we describe the algorithms we employed for measuring the similarity between these ASTs (Section~\ref{SimilarityMeasures}). Finally, we describe the search-based algorithms we used to minimize test cases using similarity measures as fitness (Section~\ref{SearchBased}).

\subsection{Test Case Code Pre-processing}
\label{preprocessing}
    
    The source code of test cases, which is in our context Java test methods, may contain information that is irrelevant to the testing rationale, such as comments and variable names, and code statements that do not exercise the system under test, such as logging statements and test oracles (assertions).
    Given that we aim to compare test cases with respect to how they exercise the system, the information above is not only irrelevant but could introduce noise in our analysis.
    Therefore, we pre-process the source code of test cases as follows.

    \begin{itemize}
        \item We removed test case names from method declarations, since they are typically different among test cases.
        \item We removed Javadoc, single- and multi-line comments, since they are simply used to document test case code.
        \item We removed logging or printing statements, since they are simply used to record test case execution.
        \item We removed test oracles, i.e., assertions, similar to Silva et al.~\cite{silva2020lccss}, since they do not exercise the system under test but rather focus on verifying the test case outcome for a given input. This includes all \textit{JUnit} assertion methods\footnote{\url{https://junit.org/junit4/javadoc/4.13/org/junit/Assert.html}}, including their parameters.

         \item Similar or even identical test cases can use different identifiers. Therefore, we normalized variable identifiers, rather than removing them, and retained their data types to maintain the data flow and logic of test cases. This was done by keeping track of variable and object identifiers according to their order of appearance in the test case code and then normalizing them in the form of $id\_1$, $id\_2$, and so on.
         \vspace{2pt}
    \end{itemize}

Listing~\ref{testcode_Before}~and~\ref{testcode_After} show a sample test case code before and after pre-processing, respectively.

\begin{table}[ht]
\centering
\vspace{-10pt}
\begin{tabular}{p{8.7cm}}

    \begin{lstlisting}[language=Java,caption=Test case before pre-processing,numbersep=2pt,label=testcode_Before,belowskip=-0.2\baselineskip]
/** 
  * Confirm that the equals method can distinguish all the required fields.
  */
public void testEquals(){
    DefaultTableXYDataset d1=new DefaultTableXYDataset();
    DefaultTableXYDataset d2=new DefaultTableXYDataset();
    assertTrue(d1.equals(d2));
    assertTrue(d2.equals(d1));
    d1.addSeries(createSeries1());
    assertFalse(d1.equals(d2));
    d2.addSeries(createSeries1());
    assertTrue(d1.equals(d2));
}\end{lstlisting}
\\  
\vspace{-10pt}
\begin{lstlisting}[language=Java,caption=Test case after pre-processing,numbersep=2pt,label=testcode_After,belowskip=-0.5\baselineskip]
public void test_case() {
    DefaultTableXYDataset id_1=new DefaultTableXYDataset();
    DefaultTableXYDataset id_2=new DefaultTableXYDataset();
    id_1.addSeries(createSeries1());
    id_2.addSeries(createSeries1());
}\end{lstlisting}
\end{tabular}
\vspace{-15pt}
\end{table}

\subsection{Transforming Test Case Code into AST}
\label{CodeAsAST}

    Processing test case code as natural language using text-based or token-based techniques does not capture its syntactical information~\cite{zhang2019novel}, thus making similarity measurement less accurate.
    To address this issue, we used Abstract Syntax Trees (AST) to preserve the syntactic structure of test case code~\cite{noonan1985algorithm}. To do this, we used an AST parser, provided by the Eclipse JDT library\footnote{\url{https://www.eclipse.org/jdt}}, to statically traverse any given test case code and transform it into a corresponding AST.
    Comparing the ASTs of test cases helps identify differences between them more precisely, such as differences in method calls, their number of parameters, or parameter values.
    AST is composed of labeled nodes, where the order of AST nodes is important as it represents the structure of test case code. Therefore, ASTs are considered \textit{labeled}, \textit{ordered} trees.

\subsection{Similarity Measures}
\label{SimilarityMeasures}
We used different algorithms to perform tree-based similarity measurement of test cases.
Depending on the way trees are traversed, algorithms may capture different information from each other.
Given the variety of coding conventions and practices according to which test cases are developed, a single similarity measure for test case minimization might perform inconsistently across projects.
Therefore, we employed four similarity measures, namely \textit{top-down} similarity (based on the top-down maximum ordered common subtree isomorphism algorithm), \textit{bottom-up} similarity (based on the bottom-up maximum ordered common subtree isomorphism algorithm), a \textit{combined} measure (merging the first two), and \textit{tree edit distance} (based on the standard tree edit distance algorithm).

Both top-down and bottom-up similarity measures focus on identifying the longest common branch of the code of test cases, but in two distinct ways.
Specifically, top-down similarity is structure-oriented, thus starting the analysis at the high-level structure of test case code, e.g., iterative or conditional blocks, followed by lower-level details in a step-by-step manner. Hence, structural differences in the code of test cases can significantly impact top-down similarity.
Bottom-up similarity, on the other hand, is detail-oriented, in which low-level details of the test case code, e.g., parameters to method calls, are analyzed first, thus making it less impacted by structural differences.
Given that bottom-up complements top-down, we further considered merging the information obtained by both of them into a new, combined similarity measure to capture both structural and detailed aspects of test cases.
Different from the above the similarity measures, tree edit distance focuses on scattered code differences between test cases rather than a common code branch and aims to capture all code changes that can make one test case similar to another.
We describe below each of the similarity measures and point to the book of Valiente~\cite{valiente2002algorithms} for more details.

\vspace{3pt}
\subsubsection{Top-down similarity measure}
\label{TopDownCommonSubtreeIsomorphism}
This measure is based on the top-down maximum common subtree isomorphism algorithm~\cite{valiente2002algorithms}.
Tree isomorphism determines whether the nodes of one tree has a bijective correspondence with the nodes of another tree.
Given that AST represents structured source code, in which the order and type of code statements is important, we considered labeled, ordered tree isomorphism.
A top-down maximum common subtree isomorphism between two labeled, ordered trees is obtained by traversing them simultaneously using \textit{preorder traversal}.
Preorder traversal ensures that the parent of each node is included in the subtree, since they are visited first, thus resulting in a top-down subtree.
Top-down similarity does not capture similar subtrees with different parent nodes. Typical examples include the same block of code but with a different loop, such as \texttt{\textit{for}} and \texttt{\textit{while}}, or calls to two different methods with the same parameters.

\vspace{3pt}
\subsubsection{Bottom-Up similarity measure}
\label{BottomUpCommonSubtreeIsomorphism}
This measure is based on the bottom-up maximum common subtree isomorphism algorithm~\cite{valiente2002algorithms}.
A bottom-up maximum ordered common subtree isomorphism between two labeled, ordered trees is obtained by first partitioning tree nodes into equivalence classes, and then finding nodes in both trees belonging to the same equivalence class with the largest size.
The size of an equivalence class is equal to the size of the bottom-up subtree rooted at a specific node.
Two nodes belong to the same equivalence class if and only if the bottom-up ordered subtrees rooted at them are isomorphic~\cite{valiente2002algorithms}. 
A \textit{postorder traversal} is then performed on $T_1$ and $T_2$ to partition children nodes into equivalence classes, starting with one tree to populate a dictionary of equivalence classes, followed by the second tree with the same dictionary shared.
The equivalence class of a visited node is obtained from the dictionary if its label and the equivalence classes of its children nodes already exist.
Once equivalence classes are assigned to all nodes of the two trees, the maximum bottom-up common subtree is obtained based on the nodes sharing the same equivalence class and having the largest bottom-up subtree rooted at them.
The bottom-up similarity measure can capture information that the top-down maximum common subtree algorithm might not. For example, in contrast to top-down similarity, if two test cases call two different methods with the same parameters, then the matching parameters of method calls are included in the bottom-up maximum common subtree. Also, if some children of the nodes in the bottom-up ordered subtrees do not match, then these nodes are not included as part of the common subtree.

\vspace{3pt}
\subsubsection{Combined similarity measure (top-down$+$bottom-up)}
\label{CombinedSimilarity}
Given that top-down and bottom-up common subtrees capture different and complementary information in the test case code, we also considered merging their resulting subtrees into a single, combined common subtree.
Combining top-down and bottom-up maximum common subtrees was performed by taking the union of nodes in both subtrees, while eliminating overlapping nodes. For each node of one subtree, we checked whether it is present in the other subtree by considering its label and the labels of its parent, siblings, and children nodes.
If there is a match, then an overlap is identified and thus not included as part of the combined common subtree.
The size of the combined common subtree is equal to the sum of the size of the unique nodes of the top-down and bottom-up common subtrees, where overlapping nodes are discarded.

\vspace{3pt}
\subsubsection{Tree edit distance similarity measure}
\label{tree_edit_distance}
This measure is based on the edit distance algorithm, which calculates the total number of elementary edit operations, i.e., insertion, deletion, and substitution of nodes, required to convert one tree into another tree~\cite{valiente2002algorithms}, which is commonly used for tree comparison.
To do this, a sequence of elementary edit operations are applied to one tree until the other tree is obtained.
Tree edit distance is not expected to be efficient when compared to previous similarity measures, especially for large ASTs, but is nevertheless an option.

\subsubsection{Similarity score calculation}
Similarity scores for the top-down, bottom-up, and combined similarity measures were calculated the same way, but differently from tree edit distance, as described below.

\begin{itemize}
    \item For top-down, bottom-up, and combined similarity measures, after identifying their maximum ordered common subtrees, a similarity score was calculated as follows.\vspace{-5pt}

    \begin{equation}
    Sim_{MaxCommonSubtree}(T_1, T_2) = \frac{2 \times |V_m|}{|V_1| + |V_2|}
    \end{equation}
    where $T_1$ and $T_2$ are two trees with a total number of $|V_1|$ and $|V_2|$ nodes, respectively. $V_m$ is the number of nodes included in the maximum ordered common subtree.
    
    \item For tree edit distance, the similarity score was calculated as follows.\vspace{-15pt}
    
    \begin{equation}
    Sim_{TreeEditDistance}(T_1, T_2) = \frac{|V_1| + |V_2| - d}{|V_1| + |V_2|}
    \end{equation}
    where $T_1$ and $T_2$ are two trees with a total number of $|V_1|$ and $|V_2|$ nodes, respectively, and $d$ is the number of tree edit operations.
\end{itemize}

\vspace{3pt}
\subsubsection{Similarity measurement implementation}
We used an open-source library, called \textsf{simpack}\footnote{\url{https://files.ifi.uzh.ch/ddis/oldweb/ddis/research/simpack}} for implementing the algorithms for top-down and bottom-up maximum ordered common subtree isomorphism and tree edit distance.
However, for the bottom-up maximum ordered common subtree isomorphism, we extended the library to support labeled trees~\cite{valiente2002algorithms} (available in our replication package~\cite{our_replication_package}). We did so by assigning unique integers to node labels, which are then used to assign equivalence classes for identifying the maximum bottom-up common subtree of two labeled, ordered trees.

\subsection{Search-based Test Case Minimization}
\label{SearchBased}
Considering that test case minimization is an NP-hard problem, we employed meta-heuristic search algorithms to help find near-optimal, feasible solutions for this problem. Meta-heuristic techniques enable us to efficiently explore the search space of minimized test suites.
Given that \ourapproach~relies on test case similarity, a search algorithm can help identify and eliminate most redundant test cases, thus producing a more diverse subset of the test suite for a given budget.
We employed two search algorithms: Genetic Algorithm (GA) and its multi-objective counterpart, namely Non-Dominated Sorting Genetic Algorithm II (NSGA-II)~\cite{deb2002fast}.

\vspace{3pt}
\subsubsection{Genetic Algorithm (GA)}
\label{GA}
This is the most widely used search algorithm in search-based software testing, inspired by evolutionary theory. Applying GA requires to properly define (a) individuals or chromosomes referring to the possible solutions, which are minimized test suites in our context, and (b) a fitness or objective function to assess the quality of each individual with respect to an optimization problem, which is test case minimization in our context, where test cases with higher similarity are eliminated.
To tailor GA to our problem, we re-formulated the optimization problem as a fixed-size subset selection problem. To do this, we represented a test suite as a binary vector whose length equals the total number of test cases before minimization, where $1$ means a test case is included and $0$ means otherwise. We selected the following three genetic operators:
(1) \textit{selection}, for which we used a binary tournament selection to select the test subset with the lower fitness among two subsets;
(2) \textit{crossover}, for which we used a customized crossover operator~\cite{pymoo} that keeps only test cases belonging to both parent test subsets, then proceed with the remaining test cases, and repeats this process until a certain number of test cases is reached; and
(3) \textit{mutation}, for which we used a permutation inversion operator to randomly select a segment of an individual and reverse its order~\cite{pymoo} to ensure fixed-size offspring.
The parameters used for our GA are consistent with what is recommended in published guidelines\footnote{\url{https://www.obitko.com/tutorials/genetic-algorithms/recommendations.php}}. Specifically, we used a population size of $100$, a mutation rate of $0.01$, and a crossover rate of $0.90$.
The GA evolution process is repeated until a termination criterion is satisfied, which is in our case when the fitness value improves by less than $0.0025$ with a minimum of $30$ generations. Once the termination criterion is reached, minimization stops and the final minimized test suite is expected to contain diverse test cases.
\vspace{3pt}
\subsubsection{Non-Dominated Sorting Genetic Algorithm II (NSGA-II)}
This is a multi-objective alternative of GA that is based on the \textit{Pareto} dominance theory~\cite{deb2002fast}. We used NSGA-II to consider two similarity measures at once in our test subset selection, where each objective consists in minimizing a distinct similarity measure. 
Individual $A$ \textit{Pareto} dominates individual $B$ if individual $A$ is at least as good as individual $B$ in all objectives, and superior to individual $B$ in at least one objective~\cite{Luke2013Metaheuristics}.
To do this, all individuals are sorted into several \textit{Pareto} non-dominated fronts and a \textit{Pareto} front rank is assigned to each individual.
We used a binary tournament selection operator, which selects test cases based on (a) the \textit{Pareto} non-domination front ranks of individuals, and (b) the crowding distance measuring the density of individuals around a particular individual. When two individuals have the same \textit{Pareto} front rank, the one with the highest crowding distance is selected. We used the same operators and termination criterion as for GA.
We considered two alternative pairs of similarity measures as fitness: 
(1) top-down \& bottom-up, to determine whether considering them as independent fitness functions leads to better results than combining their subtrees into a single similarity score (combined), and
(2) combined \& tree edit distance to assess whether the latter complements the longest common subtrees in identifying test case similarity.

\section{Validation}
\label{Validation}
This section reports on the experiments we conducted to evaluate the effectiveness and efficiency of \ourapproach. As mentioned in the introduction, the focus here is on Java projects given the unsatisfactory results obtained by previous work, as discussed in related work. We discuss below the research questions we addressed, the experimental design and dataset, and the results achieved with their practical implications.

\subsection{Research Questions}

    \noindent\textbf{\textit{RQ1: How does \ourapproach~perform under different configurations in terms of test case minimization?}}
    \vspace{3pt}
    
    \noindent Test case minimization aims to remove redundant test cases with a minimal fault detection loss within practically reasonable time. However, its performance can be influenced by the similarity measure used to compare test cases to each other. In this RQ, we assess the performance, in terms of both effectiveness and efficiency, of \ourapproach~under various configurations, each with a different combination of similarity measures, either individually (using GA) or together (using NSGA-II). We evaluated the alternative \ourapproach~configurations on each Java project using three minimization budgets ($25\%$, $50\%$, and $75\%$) widely covering the minimization range. Our choice of minimization budgets was constrained by both the very large computation time of our experiments and the fact that test engineers, according to our discussion with industry partners, usually want to preserve significant fault detection power, knowing that prioritization and selection techniques can ultimately be used to further reduce testing time. In addition, we analyzed the trade-off between effectiveness and efficiency of the alternative \ourapproach~configurations. Specifically, we addressed the following sub-RQs.

    \begin{itemize}
        \item \textbf{\textit{RQ1.1: How effectively can \ourapproach~minimize test cases?}}
        
        A similarity measure determines what information about test cases should be used to compare them. However, such information varies widely from one similarity measure to another, which can in turn affect the fault detection capability of a test case minimization technique. In this RQ, we assess the effectiveness of \ourapproach~in terms of fault detection capability for the tree minimization budgets.
        
        \vspace{4pt}
        \item \textbf{\textit{RQ1.2: How efficiently can \ourapproach~minimize test cases?}}
        
        Test case minimization should be practically scalable to projects with large test suites. However, given that similarity algorithms traverse test case ASTs differently, their execution time can vary, thus affecting the overall time required to minimize test cases. In this RQ, we assess the efficiency of \ourapproach~in terms of preparation time (taken for transforming the code of test cases into ASTs and calculating similarity scores) and minimization time (taken for running search algorithms).

    \end{itemize}
    
    \vspace{3pt}
    \noindent\textbf{\textit{RQ2: How does \ourapproach~compare to state-of-the-art black-box test case minimization techniques?}}
    \vspace{2pt}
    
    \noindent Selecting test cases arbitrarily can indeed reduce test suites, but is not a viable option as it does not take test case similarity and fault detection capability into consideration.
    To address this issue, many test case minimization techniques~\cite{khan2018systematic}, both white-box and black-box, were proposed to remove redundant test cases that are likely to detect the same faults. However, white-box techniques rely on production code, which is not always (entirely) accessible or available to test engineers and entails scalability and practicality issues in many contexts. Further, as described in Section~\ref{RelatedWork}, despite the high efficiency of some existing black-box techniques, their fault detection loss was observed to be relatively high for Java test cases, thus making them ineffective in practice.
    In this RQ, we assess the performance of \ourapproach~compared to two baseline techniques: (a) random test case minimization, a standard baseline, and (b) \fastr, a set of of novel black-box test case minimization techniques, whose efficiency was shown to be much higher than white-box techniques while achieving a comparable low fault detection capability for Java test cases.
    Further, given that both efficiency and effectiveness are important factors in selecting minimization techniques in practice, we discuss the trade-offs between fault detection rate and execution time.

\subsection{Experimental Design and Dataset}
We performed a series of experiments to evaluate the performance of \ourapproach~and assess the most effective and efficient configurations across various combinations of similarity measures and search algorithms, and compared the best configuration to baseline techniques.
Each technique was applied to each project version, independently, and was thus run $6,610$ times ($661$ projects' versions $\times$ $10$ runs). We considered all projects' versions to increase the number of instances in our experimental evaluation. However, in practice, minimization is expected to be applied only when required, i.e., many new test cases are created.
All experiments were performed on a cluster of $1,304$ nodes with $80,912$ available CPU cores, each with a $2x$ AMD Rome 7532 with 2.40 GHz CPU, 256M cache L3, 249GB RAM, running CentOS 7. Overall, our experiments took approximately three months of calendar time and $23$ years of computation time.
Moreover, to mitigate randomness in the obtained results, we ran each experiment $10$ times, each with a different random number generator seed, ranging from $1$ to $10$, to enable the reproduction of our results. The reported results are summarized for all runs using descriptive statistics.

\vspace{3pt}
\subsubsection{Minimization budgets}
For experimental purposes, the minimization budgets were set at $25\%$, $50\%$, and $75\%$ of the test suites, for the reasons mentioned above.
We focus our analysis and discussion on the results obtained for the $50\%$ minimization budget as such a percentage is sufficiently large to challenge the minimization algorithms and to show the practical significance of \ourapproach~compared to other techniques.
Results for the other two minimization budgets were consistent in terms of observations and conclusions, and can be found in our replication package~\cite{our_replication_package}.

\vspace{3pt}
\subsubsection{Baseline techniques}
~

    \vspace{3pt}
    \noindent\textbf{\textit{Random minimization.}} We used random minimization of test cases as a standard baseline in our evaluation. It is considered the simplest technique and is commonly used as a baseline of comparison for more sophisticated search algorithms~\cite{ali2009systematic}. It randomly generates subsets of test cases for any given minimization budget. Similar to other techniques, we ran random minimization $10$ times, using fixed seeds ranging from $1$ to $10$ to enable the reproduction of results. We then calculated the average fault detection rate (\textit{FDR}) across projects' versions.

    \vspace{3pt}
    \noindent\textbf{\textit{\fastr.}}
    We compared \ourapproach~to \fastr, a set of four novel black-box test case minimization techniques proposed by Cruciani et al.~\cite{cruciani2019scalable}, namely FAST++, FAST-CS, FAST-pw, and FAST-all.
    All \fastr~alternatives rely solely on the code of test cases.
    FAST++ and FAST-CS convert test case code into vectors using term frequency~\cite{turney2010frequency}, and based on these vectors, test cases are then clustered using $k$-means++~\cite{arthur2006k} (FAST++) and constructed coresets~\cite{bachem2018scalable} (FAST-CS).
    FAST-pw and FAST-all, however, use minhashing and locality-sensitive hashing~\cite{rajaraman2011mining} to identify diverse test cases based on Jaccard distance and random sampling, respectively.

    While \fastr~performed very efficiently, it achieved relatively low median \textit{FDR} results when evaluated on Java projects, ranging from $0\%$ to $22\%$ for minimization budgets from $1\%$ to $30\%$. This motivated us to compare the performance of \ourapproach~to \fastr~on test cases collected from a larger set of Java projects with many versions. Though \ourapproach~uses finer-grained information from test case code, which is likely to require longer time to execute, we aim to investigate whether it can achieve significantly higher \textit{FDR} results within practical execution time.
    We compared \ourapproach~to \fastr~using the three minimization budgets indicated above and the same evaluation procedures and metrics. 
    We relied on the publicly available implementation of \fastr\footnote{\url{https://github.com/ICSE19-FAST-R/FAST-R}} provided by its authors.
    Though \fastr~was originally evaluated on Java test classes, it could easily be adapted to Java test methods, since it takes as input (a) test code, regardless of its granularity, and (b) a mapping of faults and test cases, the former being easily mapped to test methods rather than test classes.

    Note that we did not compare with white-box techniques, since analyzing code coverage for all test cases of all project versions would be computationally challenging and is unnecessary given that the \textit{FDR} results of \fastr~have been shown to be comparable to white-box techniques.
    
\vspace{3pt}
\subsubsection{Dataset}
We evaluated \ourapproach~compared to the baseline techniques on $16$ Java projects collected from a public dataset, called \textsc{Defects4J}\footnote{\url{https://github.com/rjust/Defects4J}}, the same source of Java projects used to evaluate \fastr. Only one project from \textsc{Defects4J} was left out as it was far too large to consider for running our experiments, which already undertook months of computations.
While \fastr~was evaluated on five Java and five C projects, each with a single version, it achieved relatively lower fault detection rates on Java projects compared to C projects, where Java test cases are test classes and C test cases are command lines. 
Therefore, in this paper, we focused the implementation of \ourapproach~on Java projects and assessed it on a much larger dataset of Java projects and versions.
Each project has multiple (faulty) versions, ranging from $4$~to~$112$, with many test cases each, ranging from $152$~to~$3,916$.
Each project version contains a single \textit{real} fault associated with one or more test case failures, and was fixed by modifying the production code. We acknowledge that minimization should ideally be evaluated on project versions with multiple faults to achieve higher realism, but there exist no such public datasets with \textit{real} faults. Further, with multiple faults per version, some faults may mask other faults and it becomes very hard to determine which test case detects which fault, thus making experiments rather complicated.
Though our dataset is much larger in terms of systems and test cases than any previous test case minimization experiment, we fully realize that industrial systems can be much larger. They would, however, be unusable in our experiment as they would take far too much time. The scalability issue is further discussed in Section~\ref{Discussion}.
It is therefore easy to determine whether a test suite detects a particular fault in a given version: at least one test case fails. 
We used this dataset to evaluate \ourapproach~and baseline techniques.

Different from the \fastr's original study, we performed our evaluation on all faulty project versions. In addition, the evaluation of \fastr~on Java projects was performed at the class level where test cases are Java test classes, each of which group test methods exercising similar functionalities. This makes it impossible to identify redundant test methods within the same test class. Further, removing a whole test class can be misleading as one fault may be detected by only a subset of test methods in the test class. Thus test case minimization at the method level helps remove unnecessary test cases in a more precise manner~\cite{vasic2017file,gligoric2015practical}, which has been shown to achieve better results than those at the class level~\cite{zhang2013bridging,mei2012static,de2017test}.
Therefore, our evaluation of \ourapproach~is finer-grained as each test case is considered to be a Java test method.

We extracted the source code of test methods for each version of the projects and mapped each fault to its corresponding failing test method(s). 
Then, we transformed the test case code into ASTs, which were saved in XML format. After that, we calculated the similarity scores for each pair of test cases using the similarity measures described in Section~\ref{SimilarityMeasures}.
To reduce the time required for similarity calculation in each version during our computationally intensive experiments, we calculated the similarity scores for all test cases in the first version of each project. Then, for subsequent versions, we calculated similarity scores for only test cases in changed or newly added test files, whereas similarity scores for unchanged ones were obtained from previous versions. 

\vspace{2pt}
\subsubsection{Evaluation metrics}
~

    \vspace{1pt}
    \noindent\textbf{\textit{Fault Detection Rate (\textit{FDR}).}} Test case minimization aims to remove redundant test cases for a given budget while maintaining high \textit{FDR}. Therefore, we used \textit{FDR} to assess the effectiveness of \ourapproach. \textit{FDR} was calculated for each project as follows.\vspace{-14pt}

    \begin{equation}
    FDR = \frac{\sum_{i = 1}^{m}f_i}{m}
    \end{equation}
    where $m$ refers to the total number of versions (system faults). For each version $i$, $f_i$ equals to $1$ if at least one failing (or fault-triggering) test case is included in the minimized test suite, or~$0$~otherwise.

    \vspace{2pt}
    \noindent\textbf{\textit{Fisher's exact test.}} We used Fisher's exact test~\cite{raymond1995exact}, a non-parametric statistical hypothesis test, to assess how significant is the difference in proportions of detected faults between the alternative \ourapproach~configurations.

    \vspace{2pt}
    \noindent\textbf{\textit{Odds ratio.}} We used the odds ratio~\cite{arcuri2014hitchhiker} as an effect size measure of the magnitude of improvement of one \ourapproach~configuration over another.
    An odds ratio of $1$ indicates no difference between two techniques, whereas an odds ratio of~$>1$~indicates a higher chance for one technique to perform better than the other.

    \vspace{2pt}
    \noindent\textbf{\textit{Execution time.}} Execution time has significant practical implications for large systems and test suites. Therefore, we assessed the efficiency of \ourapproach~by computing (1) the \textit{preparation time}, taken to transform the code of test cases into ASTs and calculate similarity between all pairs of test cases, and (2) the \textit{minimization time}, taken to run search algorithms. Note that, when reporting execution time results, we did not consider the savings in execution time that can be achieved by only calculating similarity between new pairs of test cases but rather accounted for all test cases in each project version independently from other versions. This, of course, makes \ourapproach~look worse in terms of execution time.
    For each \ourapproach~configuration, we computed the average execution time for each project version.
    We also computed the execution times taken by baseline techniques.

\vspace{3pt}
\subsubsection{Similarity measures as fitness}
Our fitness functions add up similarity scores for all test case pairs, normalize the summation by the number of test case pairs in a  $[0-1]$ range, and thus quantify how similar overall test cases are in a given test suite regardless of its size.
However, we could consider a test case to be redundant if it is highly similar to at least one other test case, thus making it unnecessary to consider the other test cases. Therefore, taking the pair with the maximum similarity score
for each test case as a fitness value could better distinguish highly redundant test cases.
Moreover, in a similar vein, we could consider that only very high similarity scores truly matter in terms of test cases being redundant and that simply summing up similarity scores among pairs is not the best measurement for minimization.
Therefore, we could give more weight to test cases with higher similarity scores by squaring or exponentiating such scores.
For example, though $0.9$ and $0.8$ scores have the same difference ($0.1$) as $0.4$ and $0.3$ scores, their relative difference significantly increases after taking their squares ($0.17$ vs. $0.07$) or exponentials ($0.23$ vs. $0.14$).
As a result, removing a test case with a higher similarity score leads to a greater reduction in fitness.
Finally, for the reasons invoked above, we could also take into account, for each test case, only the pair with the maximum squared or exponentiated similarity score, thus focusing on highly redundant test cases.
In our experiments, we considered all the above alternatives for fitness and results showed that using the maximum of squared similarity scores, shown below, achieves the highest \textit{FDR}.\newpage

\setlength{\abovedisplayskip}{0pt}
\begin{equation}
   Fitness = \frac{\sum_{i,t_i\in M_n}Max_{i,j, t_i,t_j\in M_n, i \neq j}\ Sim(t_i,t_j)^2}{n}
\end{equation}
where $M_n$ is a minimized test suite of $n$ test cases, and $Sim(t_i,t_j)$ is the similarity score for each pair of test cases $t_i$ and $t_j$.

\subsection{Results}
We focus our discussion on the results achieved using the maximum of squared similarity scores as fitness for the $50\%$ minimization budget (results for other budgets lead to identical conclusions and are available in our replication package~\cite{our_replication_package}).

\begin{table*}
 \centering
 \caption{Descriptive statistics of \textit{FDR} and total execution time (in minutes) of \ourapproach~across project versions for the $50\%$ minimization budget.
 The highest \textit{FDR} and shortest execution time are highlighted in bold}
 \vspace{-6pt}
 \resizebox{1\textwidth}{!}{
 \begin{tabular}{l|>{\columncolor[gray]{0.8}}cr|>{\columncolor[gray]{0.8}}cr|>{\columncolor[gray]{0.8}}cr|>{\columncolor[gray]{0.8}}cr|>{\columncolor[gray]{0.8}}cr|>{\columncolor[gray]{0.8}}cr}

  \hline
 \multirow{1}{*}{\diagbox[height=3.65\line]{\raisebox{2ex}{Statistic}}{\raisebox{-2.3ex}{Technique}}} &  \multicolumn{8}{c}{GA} & \multicolumn{4}{|c}{NSGA-II}\\\cline{2-9}\cline{9-13}
            & \multicolumn{2}{c}{\multirow{2}{*}{Top-Down}}  & \multicolumn{2}{|c}{\multirow{2}{*}{Bottom-Up}}  & \multicolumn{2}{|c}{\multirow{2}{*}{Combined}} & \multicolumn{2}{|c}{Tree Edit}  & \multicolumn{2}{|c}{Top-Down \&}    & \multicolumn{2}{|c}{Combined \&}    \\[-3pt]
            & \multicolumn{2}{c}{} & \multicolumn{2}{|c}{} & \multicolumn{2}{|c}{}  & \multicolumn{2}{|c}{Distance}  & \multicolumn{2}{|c}{Bottom-Up}    & \multicolumn{2}{|c}{Tree Edit Distance}    \\
            \cline{2-13}
            & \textit{~FDR~} & \textit{Time~} & \textit{~FDR~} & \textit{Time~} & \textit{~FDR~} & \textit{Time~} & \textit{~FDR~} & \textit{Time~}  & \textit{~FDR~} & \textit{Time~~} & \textit{~FDR~} & \textit{Time~~~} \\
  \hline
    Min            & 0.53 &  0.42  & 0.56 &  \textbf{0.28} & 0.58          &  0.46           & \textbf{0.70} & 0.66   & 0.60          &    1.11 & \textbf{0.70} &   1.38  \\
    25\% Quantile  & 0.75 &  1.95  & 0.68 &  \textbf{1.33} & 0.75          &  2.18           & 0.76          & 3.79   & 0.74          &    5.67 & \textbf{0.78} &   7.12  \\
    Mean           & 0.78 & 70.87  & 0.74 & \textbf{67.05} & 0.80          & 72.75           & 0.81 & 82.23  & 0.78          &  235.41 & \textbf{0.82}          & 258.44  \\
    Median         & 0.79 & 12.53  & 0.72 &  \textbf{7.76} & 0.79          & 13.86           & \textbf{0.82} & 16.61  & 0.79          &   37.01 & \textbf{0.82}          &  37.15  \\
    75\% Quantile  & 0.84 & 57.88  & 0.81 & \textbf{40.27} & \textbf{0.88} & 63.01           & \textbf{0.88} & 77.77  & 0.82          &  187.74 & \textbf{0.88} &  208.38 \\
    Max            & 0.93 & 642.31 & 0.90 &         706.31 & \textbf{0.97} & 641.42 & 0.93          & \textbf{491.52} & \textbf{0.97} & 2295.20 & 0.92          & 2384.56 \\
  \hline
 \end{tabular}
 }
\vspace{-3pt}
\label{tab:FDR_Time}
\end{table*}

\vspace{2pt}
\subsubsection{RQ1 results}
Table~\ref{tab:FDR_Time} reports the \textit{FDR} and total execution time (in minutes) for \ourapproach~using GA and NSGA-II using the four similarity measures, individually and combined, for the $50\%$ minimization budget.

\vspace{2pt}
\noindent\textbf{\textit{RQ1.1 results.}}
    Table~\ref{tab:FDR_Time} shows that all \ourapproach~configurations achieved high \textit{FDR} results (mean $\geq0.74$ and median $\geq0.72$ across projects, for a $50\%$ minimization budget).
    The highest average \textit{FDR} was achieved by NSGA-II with combined \& tree edit distance (mean and median~=~$0.82$), and ranging from $0.70$ to $0.92$ across projects.
    The difference in \textit{FDR} between projects needs further investigation, as it may be attributed to variability in numbers of faults, test suite sizes, or test coding conventions, tentative explanations that remain to be confirmed.
    Overall, detecting over $80\%$ of faults when executing $50\%$ of test cases is encouraging and, as discussed below, significantly outperforms baseline techniques.
    
    Moreover, \ourapproach~using GA achieved a higher \textit{FDR} with top-down ($+0.04$ on average)
    than bottom-up across projects' versions, except for the $JacksonXml$ project where the former achieved $0.25$ lower average \textit{FDR} than the latter. One possible explanation is that, unlike other projects, failing test cases in $JacksonXml$ have higher maximum top-down similarity scores than other test cases, which indicates high redundancy among them, thus leading to the removal of some of them. 
    
    Our results also show that \ourapproach~using GA with combined similarity yielded an even higher \textit{FDR} (mean~=~$0.80$, median~=~$0.79$) than with top-down, only $0.01$ and $0.02$ lower than GA with tree edit distance and NSGA-II with combined \& tree edit distance, respectively. This suggests that taking the union of the top-down and bottom-up common subtrees helped capture additional, relevant information about test case commonalities. Further, \ourapproach~using GA with combined similarity achieved a higher \textit{FDR} than NSGA-II with top-down \& bottom-up ($0.78$), thus suggesting that a simpler minimization algorithm (GA) with a single similarity measure (combined similarity) outperforms a more sophisticated and expensive algorithm (NSGA-II) relying on two similarity measures.
    
    Fisher's exact test results revealed no significant \textit{FDR} differences between \ourapproach~using GA with combined similarity and both GA with tree edit distance and NSGA-II with combined \& tree edit distance ($p-value>0.05$). More details about the results of Fisher's exact test can be found in our replication package~\cite{our_replication_package}.
    In other words, even though GA with combined similarity yielded a slightly lower ($0.01-0.02$ less) \textit{FDR} on average, there is no evidence that this difference is statistically significant as results vary across projects' versions. Compared to GA with tree edit distance, GA with combined similarity achieved a higher \textit{FDR} for four projects ($Collections$, $Csv$, $JacksonDatabind$, and $Time$) and the same \textit{FDR} for four projects ($Cli$, $Compress$, $Jsoup$, and $JxPath$).

    Overall, our results suggest that, when accounting only for \textit{FDR}, \ourapproach~alternatives using either GA or NSGA-II, with tree edit distance and/or combined similarity as fitness, are roughly equivalent for all minimization budgets.

\vspace{2pt}
\noindent\textbf{\textit{RQ1.2 results.}}
   Execution time results in Table~\ref{tab:FDR_Time} combines both preparation time and minimization time for each project version (see detailed results in our replication package~\cite{our_replication_package}).
   We observe that the average execution time for the whole \ourapproach~process using GA, with all configurations, ranges from $1.1$ to $1.4$ hours on average  per project version (with a much lower median of $7.8-16.6$ minutes due to two relatively larger projects). Given the application context of test case minimization, where redundant test cases are removed on an occasional basis when many new test cases are created for major releases, such execution times are acceptable in practice, as further discussed below. However, the execution time of \ourapproach~using NSGA-II was nearly three times longer than that of GA (mean~=~$3.9-4.3$ hours and median~=~$37$ minutes).
   This result suggests that \ourapproach~using NSGA-II with two similarity measures fares much worse in terms of execution time, while offering no significant improvement in terms of \textit{FDR}. 
   
    \vspace{2pt}
    \textbf{\textit{Preparation Time.}}
    We found that the time \ourapproach~took for transforming test case code into ASTs was negligible ($<1\%$ of total time), whereas the time for calculating similarity scores was much longer.
    Specifically, the average execution time for calculating top-down, bottom-up, and combined similarity scores for each pair of test cases across project versions was $3.42e^{-06}s$, $4.93e^{-05}s$, and $1.75e^{-04}s$, respectively. 
    In sharp contrast, though achieving the highest \textit{FDR} when using GA, we found that tree edit distance, as expected, took at least an order of magnitude longer to calculate than other similarity measures (an average of $35$ minutes per each project version, compared to $3$ minutes for combined similarity).
    Therefore, on larger projects with much larger test suites and ASTs, such as those commonly found in the industry, the absolute computation time difference between tree edit distance and other similarity measures is expected to be large and probably crucial in terms of applicability.
    Hence, in terms of preparation time, \ourapproach~with combined similarity is more scalable and a better alternative than tree edit distance in practice.
   
   \vspace{2pt}
   \textbf{\textit{Minimization Time.}}
   We found that the time \ourapproach~took for minimizing test cases using NSGA-II with both top-down \& bottom-up and with combined \& tree edit distance (mean~=~$3.7-3.9$ hours and median~=~$31-37$ minutes) was about three times longer than that of GA with combined similarity.
   In addition, we found that, though took much longer to calculate similarity, tree edit distance took less minimization time, since it converged faster, than combined similarity using GA. However, when considering the execution time for the whole process, \ourapproach~ran faster when using GA with combined similarity for all projects, except for $Time$. 
   As a result, \ourapproach~using GA with combined similarity is more scalable than with tree edit distance and far more efficient than using NSGA-II, while achieving comparable \textit{FDR} results, thus making it the best configuration. This can be particularly important on large projects and test suites. 

\summarybox{\textbf{RQ1 summary.} 
    \ourapproach~achieved high \textit{FDR} results ($0.82$ on average) and ran within practically acceptable time ($1.1-4.3$ hours on average) when running $50\%$ of test cases, with combined similarity using GA being the best configuration when considering both effectiveness ($0.80$ \textit{FDR} on average) and efficiency ($1.2$ hours on average). Results were consistent for other minimization budgets ($25\%$ and $75\%$).
}

\vspace{-2pt}
\subsubsection{RQ2 results}
Table~\ref{tab:results_with_baselines} compares, in terms of \textit{FDR} and total execution time for the $50\%$ minimization budget, the best \ourapproach~configuration (GA with combined similarity) to the baseline techniques: \fastr~(FAST++, FAST-CS, FAST-pw, and FAST-all) and random minimization.

\begin{table*}
 \centering
 \caption{Descriptive statistics of \textit{FDR} and total execution time (in seconds) for the $50\%$ minimization budget, across projects' versions, of \ourapproach~using GA with combined similarity compared to \fastr~and random minimization. The highest \textit{FDR} and shortest execution time are highlighted in bold}
 \vspace{-1.5pt}
 \resizebox{1\textwidth}{!}{
 \begin{tabular}{l|>{\columncolor[gray]{0.8}}cr|>{\columncolor[gray]{0.8}}cr|>{\columncolor[gray]{0.8}}cr|>{\columncolor[gray]{0.8}}cr|>{\columncolor[gray]{0.8}}cr|>{\columncolor[gray]{0.8}}cr}

  \hline
          \multirow{1}{*}{\diagbox[height=2.65\line]{\raisebox{1.3ex}{Statistic}}{\raisebox{-1.2ex}{Technique}}} & \multicolumn{2}{c}{\ourapproach}  & \multicolumn{2}{|c}{\multirow{2}{*}{FAST++}}  & \multicolumn{2}{|c}{\multirow{2}{*}{FAST-CS}} & \multicolumn{2}{|c}{\multirow{2}{*}{FAST-pw}} &  \multicolumn{2}{|c}{\multirow{2}{*}{FAST-all}} & \multicolumn{2}{|c}{Random}    \\[-3pt]
            & \multicolumn{2}{c}{GA/Combined} & \multicolumn{2}{|c}{} & \multicolumn{2}{|c}{}  & \multicolumn{2}{|c}{}  & \multicolumn{2}{|c}{}    & \multicolumn{2}{|c}{minimization}    \\
            \cline{2-13}
            & \textit{~FDR~} & \textit{Time~} & \textit{~FDR~} & \textit{Time~} & \textit{~FDR~} & \textit{Time~} & \textit{~FDR~} & \textit{Time~}  & \textit{~FDR~} & \textit{Time~~} & \textit{~FDR~} & \textit{Time~~~} \\
  \hline
    Min            & \textbf{0.58} &     27.58  &  0.51  &  0.06  &  0.48  &  0.06  &  0.25  &  0.45  &  0.38  & 0.42  &  0.17  &  \textbf{0.0012}  \\
    25\% Quantile  & \textbf{0.75} &    130.83  &  0.57  &  0.10           &  0.57  &  0.08  &  0.38  &  0.97  &  0.54  & 0.90  &  0.44  &  \textbf{0.0013}  \\
    Mean           & \textbf{0.80} &  4,364.76  &  0.61  &  0.44           &  0.60  &  0.20  &  0.47  &  4.14  &  0.59  & 2.78  &  0.52  &  \textbf{0.0021}  \\
    Median         & \textbf{0.79} &    831.71  &  0.60  &  0.19           &  0.60  &  0.14  &  0.47  &  2.21  &  0.62  & 1.82  &  0.50  &  \textbf{0.0017}  \\
    75\% Quantile  & \textbf{0.88} &   3,780.37 &  0.65  &  0.63           &  0.64  &  0.29  &  0.54  &  6.37  &  0.66  & 4.27  &  0.57  &  \textbf{0.0025}  \\
    Max            & \textbf{0.97} &  38,485.15 &  0.72  &  2.17           &  0.71  &  0.63  &  0.72  &  17.11 &  0.70  & 9.10  &  1.00  &  \textbf{0.0056}  \\
    
  \hline
 \end{tabular}
 }
\label{tab:results_with_baselines}
\vspace{-1pt}
\end{table*}

    \vspace{2pt}
    \textbf{\textit{FDR} results.}
    We observe that \ourapproach~using GA with combined similarity systematically outperformed random minimization with a significantly higher \textit{FDR} ($+0.28$ on average).
    In addition, all \fastr~alternatives, except FAST-pw, outperformed random minimization, with FAST++ being the best \fastr~alternative ($0.61$ on average), followed by FAST-CS ($0.60$ on average).
    However, compared to \ourapproach~(GA with combined similarity), all \fastr~alternatives achieved significantly lower \textit{FDR} results across projects, with an average \textit{FDR} difference of $-0.19$ between FAST++ and \ourapproach, thus making \ourapproach~much more effective in practice.

    \vspace{2pt}
    \textbf{Execution time results.}
    We observe that all \fastr~alternatives ran much faster than~\ourapproach, in terms of both preparation and minimization, with an average total execution time of $0.20-4.14$ seconds across projects' versions (FAST-CS was the fastest technique, with $0.20$ seconds). 
    However, given its considerably low \textit{FDR}, i.e., missing about $40\%$ of faults when executing $50\%$ of test cases, \fastr~is not a practically viable option in many contexts.\footnote{As a side note, though out of the scope of this work, \fastr~achieved lower average \textit{FDR} results (up to $0.56$, achieved by FAST++) and took slightly longer ($0.23-5.75$ seconds) when evaluated at the class level (Java test classes), as in the original \fastr~study.}
    Test case minimization is typically performed on an occasional basis~\cite{noemmer2020evaluation}, usually at certain milestones, such as new major releases when many new test cases are created. Therefore, given that \ourapproach~achieves much higher \textit{FDR} and runs within practically acceptable (though longer) time, it is the most advantageous choice in many practical contexts.\vspace{3pt}

\summarybox{\textbf{RQ2 summary.}
    \ourapproach~outperformed baseline techniques by achieving significantly higher \textit{FDR} results than \fastr~($+0.19$ on average) and random minimization ($+0.28$ on average), while running within practically acceptable (though longer) time ($1.2$ hours on average) given the application context.
}

\subsection{Discussion}
\label{Discussion}
\vspace{2pt}
\noindent \textbf{\textit{Effective test case minimization with easily accessible information.}}
Our results showed that \ourapproach~performs significantly better than baseline techniques, thus enabling test engineers to run test suites for a desired budget while being more likely to maintain an acceptable \textit{FDR}. This is done without requiring production code analysis, a significant practical advantage in many contexts.
While \ourapproach~achieved high \textit{FDR} result using similarity measures, both individually and combined, there is still room for improvement in terms of \textit{FDR}, which could be achieved by considering additional similarity measures, thus capturing various and complementary aspects relevant to test case commonalities.
Also, similarity measurement in \ourapproach~considered a variety of potentially relevant information in test case code. For example, similarity between method calls considers their names, number of parameters, and parameter values. Disregarding some of these details could result in a higher similarity and better results. The importance of these details may also differ from one project to another, which suggests that test engineers might need to tailor the definition of similarity to their needs and context. 

\vspace{5pt}
\noindent \textbf{\textit{Effective versus efficient test case minimization.}}
Our results showed that GA with combined similarity is the best configuration for \ourapproach, given its effectiveness and efficiency compared to other configurations. 
Specifically, it achieved a high \textit{FDR} that is comparable to GA with tree edit distance and NSGA-II alternatives, while taking much less time to calculate. 
However, despite the considerably longer time taken to calculate tree edit distance compared to combined similarity, it took relatively less time to minimize test cases as it converges faster on all projects. Still, when accounting for the total execution time, \ourapproach~using GA with combined similarity ran faster than tree edit distance on the majority of the projects, thus making it the best configuration.
Moreover, though very efficient, random minimization and \fastr~alternatives performed significantly worse than \ourapproach~in terms \textit{FDR}. 
Given that test case minimization is typically performed on an occasional basis, such as major releases with many new test cases, even if \ourapproach~takes much longer than \fastr~due to its finer-grained similarity measures and search algorithms, it still runs in practical time ($1.2$ hours on average) and achieves much higher \textit{FDR}, thus making it a better choice in many practical contexts.

\vspace{4pt}
\noindent \textbf{\textit{Scalability.}}
On the largest project in our dataset, $Time$, which has nearly $4k$ test cases, \ourapproach~took more than $10$ hours, on average, per version, largely due to the search producing an optimal minimized test suite. Though our test case minimization experiment is the largest to date, industrial systems can be much larger than the ones in our dataset. Overall, we observed that the execution time of the search algorithm increases quadratically with the number of test cases, regardless of the similarity measure.
As a result, there is obviously a limit, in terms of system and test suite sizes, to any minimization technique. Therefore, future research should devise ways of pushing the scalability boundary of \ourapproach~further while preserving high \textit{FDR} results.
For instance, similarity and search fitness computations can be easily parallelized to significantly improve scalability. Indeed, all similarity computations of test case pairs and fitness values of minimized sets in a population are independent and can be run on different cores~\cite{cantu1998survey,katoch2021review}. Other search algorithms~\cite{Luke2013Metaheuristics} should also be investigated in the future to identify better trade-offs between efficiency and effectiveness.

\vspace{4pt}
\noindent \textbf{\textit{Application of test case minimization in practice.}}
In practice, when there is a major release with many new test cases created and a testing budget is set, similarity scores of new test case pairs are calculated. Then, search is performed to find a subset of the test suite that minimizes the similarity between test cases within the given test budget. The resulting minimized test suite is later used for regression testing in the subsequent code versions.

\section{Threats to validity}
\label{Threats}
\vspace{-0.4pt}
\noindent\textbf{Construct Validity}
Construct threats to validity are concerned with the degree to which our analyses measure what we claim to analyze.
Test cases may contain information that is irrelevant to the testing rationale, which could introduce noise when measuring similarity between test cases. To mitigate this threat, \ourapproach~pre-processed test cases and transformed their code into ASTs.

\vspace{3pt}
\noindent\textbf{Internal Validity}
Internal threats to validity are concerned with the ability to draw conclusions from our experimental results.
Compared to \fastr, \ourapproach~was evaluated on finer-grained data: test cases were considered to be Java test methods instead of test classes. However, given that \fastr~was originally evaluated based on Java test classes, its performance on test methods could be inconsistent in our experiments. To mitigate this threat, we evaluated all techniques on the same finer-grained data. We observe that \fastr's performance was consistent with what was originally reported~\cite{cruciani2019scalable}, in terms of both effectiveness and efficiency, and even worse when evaluated at the class level.
Moreover, to mitigate randomness in the obtained results, we ran each experiment $10$ times, with fixed seeds to enable the reproduction of our results.
Finally, we removed all assertions, including their parameters as we assume they do not include any method calls that exercise the system under test (with side effects) as their main goal is to verify the test case outcome. Calling methods with side effects inside assertions is not considered a good testing practice. Test cases may become similar in our context when assertions are removed, meaning they exercise the same behavior, resulting in excluding one of them. However, assertions might still contain relevant information regarding test case similarity and retaining them might increase FDR. Future research should further investigate this point.

\vspace{3pt}
\noindent\textbf{External Validity}
External threats are concerned with the ability to generalize our results.
Our evaluation was performed on a large dataset extracted from $16$ Java projects collected from \textsc{Defects4J}, including $661$ faulty versions. However, unlike \fastr, we did not evaluate \ourapproach~on C projects,
though it can a priori be applied to test cases written in other languages provided the availability of tools for transforming test case code into ASTs. Future research should assess \ourapproach's performance on test cases written in other programming languages to devise more general conclusions.

\section{Related Work}
\label{RelatedWork}
\vspace{-4pt}
Test case minimization is an NP-hard problem, hence there are many techniques that have been proposed~\cite{yoo2012regression,khan2018systematic} to find near-optimal, feasible solutions to address it.
Such techniques can be classified into three categories: greedy heuristics-based, clustering-based, and search-based.

\vspace{3pt}
\noindent\textbf{Greedy heuristics-based test case minimization.}
Miranda et~al.~\cite{miranda2017scope} used greedy heuristics~\cite{chen1970heuristics} to perform test case minimization that iteratively selects test cases based on their code coverage until all the target program entities, e.g., statements, for a given testing input are covered.
This technique achieved fault detection rates ranging from $0.52$ to $0.69$ for running $2.6\%$ to $11.3\%$ of test cases.
Noemmer et~al.~\cite{noemmer2020evaluation} also used greedy heuristics to perform test case minimization using statement coverage. Mutation testing was used to evaluate this technique, which obtained a loss in mutation scores ranging from $3.5\%$ to $21\%$ for running $7\%$ to $50\%$ of test cases.
However, these techniques require to analyze production code, i.e., white-box, and do not enable targeting specific minimization budgets, thus having scalability and applicability issues in practice.

\vspace{3pt}
\noindent\textbf{Clustering-based test case minimization.}
Cruciani et~al.~\cite{cruciani2019scalable} recently proposed a novel technique, called \fastr, to perform test case minimization by relying solely on test case code (black-box), which is converted into vectors using a term frequency model~\cite{turney2010frequency}.
Based on these vectors, clustering was used to partition test cases into clusters, where the centroids of the clusters were selected as the minimized set of test cases. It was evaluated on five Java and five C projects, each with a single version. Though it was more efficient than white-box techniques, it achieved relatively low median fault detection rates on Java projects, ranging from $0.18$ to $0.22$ for minimization budgets ranging from $1\%$ to $30\%$. 
In addition, the experiment considered Java test classes to be test cases, where test methods exercising similar functionalities are grouped together, thus making it impossible to identify redundant test methods within the same test class.
Similarly, Coviello et al.~\cite{coviello2018clustering} and Viggiato et al.~\cite{viggiato2022identifying} proposed clustering-based test cases minimization techniques relying on code coverage or test cases written in natural language, thus making them not easy to apply in practice as such information are not always accessible by test engineers.

\vspace{3pt}
\noindent\textbf{Search-based test case minimization.}
Hemmati et~al.~\cite{hemmati2013achieving} used a search algorithm that relies on test case similarity calculated using model-based features to perform test case minimization.
This technique was evaluated on two relatively small industrial systems and achieved average fault detection rates ranging from $0.60$ to $1.00$ for running $4\%$ to $14\%$ of test cases.
Similarly, Zhang et al.~\cite{zhang2019uncertainty} and Wang et al.~\cite{wang2015cost} proposed model-based test case minimization techniques relying on multi-objective search algorithms, such as NSGA-II, MOCell~\cite{nebro2009mocell}, SPEA2~\cite{zitzler2001spea2}.
However, the information required by the above techniques is not always available to test engineers.\\\\

\vspace{-1pt}
\noindent\textbf{Summary.} In contrast to the above techniques, except for \fastr, \ourapproach~relies exclusively on test case code and can help test engineers target any pre-set minimization budget, thus making it more applicable in practice. 
Compared to \fastr, \ourapproach~achieved significantly higher ($+0.19$) fault detection rates within practically acceptable (though significantly longer) execution time. Moreover, \ourapproach~was evaluated on a larger, finer-grained dataset of $16$ Java projects with $661$ faulty versions, thus making it by far the largest experiment to date for test case minimization.

\section{Conclusion}
\label{Conclusion}
In this paper, we proposed \ourapproach, a black-box test case (suite) minimization technique based on the Abstract Syntax Tree (AST) similarity of test code and evolutionary search algorithms. We investigated four tree-based similarity measures, namely top-down, bottom-up, combined (merging the first two), and tree edit distance. We employed Genetic Algorithms (GA) and its multi-objective counterpart (NSGA-II), to perform test case minimization using the above similarity measures to define alternative fitness functions.
We evaluated various configurations of \ourapproach~on a large dataset of $16$ Java projects with $661$ (faulty) versions collected from \textsc{Defects4J}. We used the Fault Detection Rate (\textit{FDR}) and execution time evaluation metrics to respectively assess the effectiveness and efficiency of \ourapproach, using three practical minimization budgets ranging from $25\%$ to $75\%$. We identified the best \ourapproach~configuration and compared it to \fastr, a recently proposed set of black-box test case minimization techniques, and random minimization, a standard baseline.
For a minimization budget of $50\%$, we observed that all \ourapproach~configurations achieved significantly higher \textit{FDR} results ($0.82$ on average) compared to \fastr~($0.61$ on average) and random minimization ($0.52$ on average).
In addition, all \ourapproach~configurations ran within practically acceptable time ($1.1-4.3$ hours on average), with combined similarity using GA being the best configuration when considering both effectiveness and efficiency ($1.2$ hours on average).
Results were consistent for other budgets (25\% and 75\%).

\vspace{4pt}
\noindent\textbf{Future work.}
We aim to extend \ourapproach~in the future to use additional similarity measures, such as Normalized Compression Distance (NCD)~\cite{feldt2016test}, and other code representation techniques using language models, such as
CodeBERT~\cite{feng2020codebert} and
TreeBERT~\cite{jiang2021treebert}. We also aim to expand our evaluation to consider projects using other programming languages and larger industrial systems.

\section{Data Availability}
\label{DataAvailability}
The replication package of our experiments, including the data, code, results for other minimization budgets, and detailed \textit{FDR} and execution time results of \ourapproach~and baseline techniques, is available on Zenodo~\cite{our_replication_package}.

\bibliographystyle{unsrt}
\bibliography{paper}
\end{document}